\documentclass[twocolumn,twoside,superscriptaddress,showpacs,showkeys]{revtex4}
\usepackage{graphicx}
\usepackage{amssymb}
\usepackage{bm}
\usepackage{dcolumn}
\begin{document}
\setcounter{page}{1}
\title[Smoothing of all-sky survey map]{Smoothing of all-sky survey map with Fisher-von Mises function}
\author{Kwang-Il \surname{Seon}}
\affiliation{Korea Astronomy and Space Science Institute, 61-1 Hwaam-dong, Yuseong-gu Daejeon 305-348}
\email{kiseon@kasi.re.kr}
\thanks{Fax: +82-42-865-2020}
\date[]{Received October 4 2005}

\begin{abstract}
Convolution of all-sky survey data with a smoothing function is crucial in
calculating the smooth surface brightness of the sky survey data.
The convolution is usually performed using the spherical version of the convolution theorem.
However, a Gaussian function, applicable only in flat-sky approximation,
has been usually adopted as a smoothing kernel.
In this paper, we present an exact analytic solution of the spherical harmonic transformation
of a Fisher-von Mises function,
the mathematically version of a Gaussian function in spherical space.
We also obtain the approximate solutions $\exp[-l(l+1)/2\kappa]$.
The exact and approximate solutions may be useful when an astrophysical survey map is convolved with a smoothing function
of $\kappa>1$.
\end{abstract}

\pacs{95.80.+p, 07.05.Kf, 98.80.Es, 02.30.Nw}
\keywords{Sky surveys, Data analysis, Observational cosmology, Fourier Analysis}
\maketitle
\section{INTRODUCTION}
In order to compare the results of theoretical calculations
with an astrophysical observation, it is necessary to convolve the theoretical calculations
with a measure of the angular response of the detector and to
calculate the
quantities which correspond to those being observed.
The angular response is often called the ``beam function''.

In particular, knowledge of the beam profile and its spherical harmonic transformation
is of critical importance for
interpreting data from cosmic microwave background (CMB) experiments.
A common simplifying assumption in the spherical data analysis
is to take the experimental beam response to have a circular Gaussian profile
as in the flat-sky approximation \cite{Gott1990,White1992}.
More recently, methods to perform the convolution of a CMB map with a general
asymmetric beam, an elliptical Gaussian beam profile have been also investigated
\cite{Fosalba2002,Tristram2004}.

It is often required to convolve spherical data, such as all-sky survey maps,
with a smoothing function.
A circular Gaussian beam function has been also used to take the angular position errors
into account in the procedure of computing the angular power spectrum of BASTSE 3B gamma-ray
bursts \cite{Tegmark1996}.
Smoothing of two-dimensional spherical data is also a procedure routinely
used in many fields of sky survey data analysis \cite{Finkbeiner2003}.

A Gaussian function in flat-sky approximation is applicable
only when the beam width $\sigma\ll 1$.
The spherical version of a Gaussian function is known as
a Fisher-von Mises function by mathematicians \cite{Fisher1987}.
Fig.~1 shows substantial differences between a Gaussian function and
a Fisher-von Mises function when the beam width $\sigma\lesssim 1$.
Thus, it is worth to investigate the convolution of spherical data
with a Fisher-von Mises distribution function.
In this paper, we investigate the spherical harmonic transformation
of a Fisher-von Mises function, which is required for the convolution of
a spherical data with a beam width $\sigma\lesssim 1$.
We found the exact solution, recursive relation, and generating function of
the spherical harmonic coefficients.

The FIMS (FUV IMaging Spectrograph) instrument was launched onboard
the first Korean Science and Technology SATellite STSAT-1
on 27 September 2003 \cite{Seon2003,Seon2005a,Seon2005b}.
Its spatial resolution is about $10'-30'$ due to the optical properties of the FIMS instrument and
errors in attitude knowledge provide by spacecraft.
The first far-ultraviolet (FUV) sky maps are in preparation to be published.
Analysis of the FUV sky survey data obtained with FIMS motivated the present work.

\section{Spherical Harmonic Convolution}
We describe the spherical harmonic convolution theorem in order to
define naming conventions.
Let the continuous function $A(\theta,\phi)$ represent a map of the sky, and
let $B(\theta)$ be the (azimuthally symmetric) smoothing kernel.
If we expand a sky map $A(\theta,\phi)$ in spherical harmonics as
\begin{equation}
A(\theta,\phi)=\sum_{l=0}^{\infty}\sum_{m=-l}^{l}A_{lm}Y_{lm}(\theta,\phi), \label{eq:sph}
\end{equation}
the coefficients $A_{lm}$ are given by
\begin{equation}
A_{lm}=\int Y^*_{lm}(\theta,\phi)A(\theta,\phi)d\Omega.
\end{equation}
Similarly, let us expand the beam function in Legendre polynomials as
\begin{eqnarray*}
B(\theta)&=&\sum_{l=0}^\infty\sqrt{\frac{2l+1}{4\pi}}B_{l0} Y_{l0}(\theta,\phi).
\end{eqnarray*}
Here, we should note the different convention for the coefficients $B_l$.
Because of azimuthal symmetry of $B$, the $B_{lm}$ coefficients vanish for $m\neq 0$
so there is no sum of $m$, and we abbreviate $B_{l0}$ with $B_l$.
Inserting the definition of $Y_{l0}$ gives
\begin{eqnarray}
B(\theta)&=&\sum_{l=0}^\infty\left(\frac{2l+1}{4\pi}\right)B_l P_l(\cos\theta). \label{eq:def}
\end{eqnarray}
The coefficients $B_{l}$ are, then, given by
\begin{equation}
B_{l}=\int_{0}^{\pi}B(\theta)P_{l}(\cos\theta)\sin\theta d\theta.
\end{equation}
The convolution of $A$ with $B$ is defined by
\begin{eqnarray}
H(\theta,\phi)=A(\theta,\phi)*B(\theta)=\int d\Omega' A(\theta',\phi')B(\gamma), \nonumber
\end{eqnarray}
where $\gamma$ is the angle between $(\theta,\phi)$ and $(\theta',\phi')$.
If we denote the spherical harmonic transform of $H(\theta,\phi)$ by $H_{lm}$ following the definition Eq.~(\ref{eq:sph}),
we have the spherical harmonic convolution theorem \cite{Tegmark1996},
\begin{equation}
H_{lm}=B_{l}A_{lm}.
\end{equation}

\begin{figure}[t!]
\includegraphics[width=8cm]{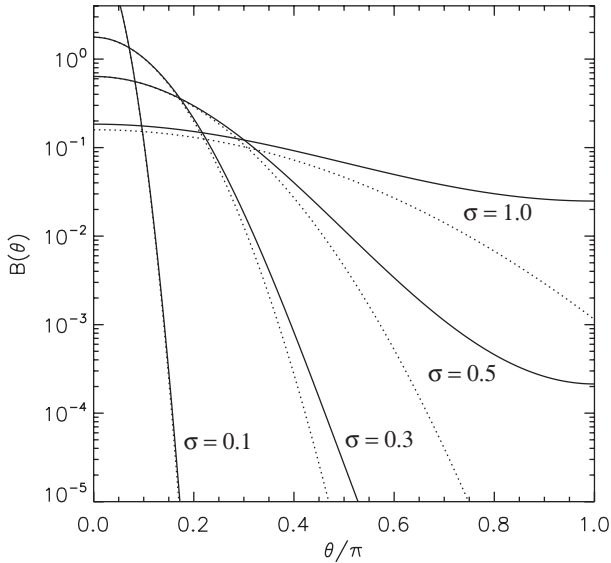}
\caption{Comparison of Gaussian and Fisher-von Mises functions with beam widths $\sigma = 0.1, 0.3, 0.5$, and 1.0.
Solid and dotted lines denote Fisher-von Mises and Gaussian functions, respectively.}
\label{fig1}
\end{figure}

\section{Spherical Harmonic Transformation}

A Fisher-von Mises function \cite{Fisher1987} is defined by
\begin{equation}
B(\theta)=\frac{\kappa\exp(\kappa\cos\theta)}{4\pi\sinh(\kappa)},
\end{equation}
characterized by the {\em concentration parameter} $\kappa$.
The larger the value of $\kappa$ the more the function is concentrated
towards the direction $\theta=0$.
The concentration parameter is related to the usual beam width by $\kappa = 1/\sigma^2$.
This is often considered by mathematicians to be the spherical version of a
Gaussian distribution, and it reduces to
\begin{equation}
B(\theta)=\frac{\exp(-\theta^2/2\sigma^2)}{2\pi\sigma^2}\label{eq:gauss}
\end{equation}
when $\sigma\ll1$ or $\kappa\gg 1$.
This is the smooth kernel which has been commonly used for the
convolution of all-sky survey maps in literatures \cite{Gott1990,White1992,Tegmark1996}.
The Fisher-von Mises function has the advantage that it is correctly normalized (its integral over
the sphere is unity) for arbitrarily large beam width $\sigma$, which is not the case
for the plane Gaussian of Eq.~(\ref{eq:gauss}).

Substituting $x=\cos\theta$ into Eq.~(\ref{eq:def}), we obtain
\begin{eqnarray}
B_l=\int_{-1}^{+1}dx P_l(x)\frac{\kappa\exp(\kappa x)}{2\sinh\kappa}.
\end{eqnarray}
Integrating this by parts, we obtain
\begin{eqnarray}
B_l&=&\frac{1}{2\sinh\kappa}\left[e^\kappa P_l(1)-e^{-\kappa}P_l(-1)
-\int_{-1}^{+1}dx e^{\kappa x}P'_{l}(x)\right]. \nonumber \\
\label{eq:parts}
\end{eqnarray}
Using the recursion relation
$P'_{l+1}(x)-P'_{l-1}(x)=(2l+1)P_l(x)$ and the properties
$P_{l}(1)=1$ and $P_{l}(-1)=(-1)^{l}$ \cite{Arfken1985},
we have the recursion relation for the
spherical harmonic coefficients $B_l$,
\begin{equation}
B_{l+1}-B_{l-1}=-(2l+1)B_{l}/\kappa. \label{eq:recursion}
\end{equation}
The first two coefficients, $B_0$ and $B_1$, can be easily obtained from Eq.~(\ref{eq:parts}) as
\begin{eqnarray*}
B_0&=&1,~~{\rm and} \\
B_1&=&\cosh\kappa - 1/\kappa.
\end{eqnarray*}
Using theses, we have also the following recursion relation.
\begin{equation}
B_{l+1}+B_{l}=\frac{e^{\kappa}}{2\sinh\kappa}-\frac{1}{\kappa}\sum_{n=0}^{l}(2n+1)B_{n}.\nonumber
\end{equation}
Now, we obtain an explicit formula for the coefficients $B_l$ by continuing the
integration by parts.
\begin{eqnarray}
B_l
&=&\frac{1}{2\sinh\kappa}\sum_{n=0}^{l}\left(-\frac{1}{\kappa}\right)^n
  \left[e^{\kappa x}P^{(n)}_{l}(1)-e^{-\kappa x}P^{(n)}_{l}(-1)\right] \nonumber
\end{eqnarray}
Here, $P^{(n)}_{l}=d^n P/dx^n$.
Using the Rodrigue's formula \cite{Arfken1985}
\begin{eqnarray*}
P_{l}(x)=\frac{1}{2^{l}l!}\frac{d^{l}}{dx^{l}}\left( x^{2}-1\right)^{l},
\end{eqnarray*}
we have $P_{l}^{(n)}(-1)=(-1)^{l+n}P_{l}^{(n)}(1)$.
Thus the formula for the coefficients $B_l$ is given by
\begin{eqnarray}
B_l&=& \frac{1}{2\sinh\kappa}\sum_{n=0}^{l}\left(-\frac{1}{\kappa}\right)^n
  \left[e^{\kappa x}_{l}-e^{-\kappa x}(-1)^{l+n}\right]P^{(n)}_{l}(1) \nonumber
\end{eqnarray}
In order to calculate $P^{(n)}_{l}(1)$, we substitute
$x^{2}-1=2(x-1)[1+(x-1)/2]$ into the equation and obtain
\begin{eqnarray*}
P_{l}^{(n)}(x) &=&\frac{1}{l!}\frac{d^{l+n}}{dx^{l+n}}(x-1)^{l}\left(1+\frac{x-1}{2}\right)^{l} \\
  &=&\frac{d^{l+n}}{dx^{l+n}}\sum_{m=0}^{l}\frac{1}{2^{m}}\frac{1}{(l-m)!m!}(x-1)^{l+m} \\
  &=&\sum_{m=n}^{l}\frac{1}{2^{m}}\frac{1}{(l-m)!m!}\frac{(l+m)!}{(m-n)!}(x-1)^{m-n}.
\end{eqnarray*}
Thus, we have
\begin{eqnarray}
P_{l}^{(n)}(1) &=&\frac{1}{2^{n}}\frac{(l+n)!}{(l-n)!n!}.
\end{eqnarray}

Further, we can rewrite the formula when $l>0$ is odd,
\begin{eqnarray*}
B_{l}&=&\coth\kappa\sum_{m=0}^{(l-1)/2}\left(\frac{1}{\kappa}\right)^{2m}P_{l}^{(2m)}(1) \\
    &&-\sum_{m=0}^{(l-1)/2}\left(\frac{1}{\kappa}\right)^{2m+1}P_{l}^{(2m+1)}(1) \\
    &=&\sum_{n=0}^{l}\left(-\frac{1}{\kappa }\right)^{n}P_{l}^{(n)}(1)\\
    &&+(\coth\kappa-1)\sum_{m=0}^{(l-1)/2}\left(\frac{1}{\kappa}\right)^{2m}P_{l}^{(2m)}(1)
\end{eqnarray*}
When $l>0$ is even,
\begin{eqnarray*}
B_{l}&=&\sum_{m=0}^{l/2}\left( \frac{1}{\kappa }\right)^{2m}P_{l}^{(2m)}(1) \\
  &&-\coth\kappa\sum_{m=0}^{l/2-1}\left(\frac{1}{\kappa}\right)^{2m+1}P_{l}^{(2m+1)}(1)\\
  &=&\sum_{n=0}^{l}\left(-\frac{1}{\kappa }\right)^{n}P_{l}^{(n)}(1)\\
    &&-(\coth\kappa-1)\sum_{m=0}^{l/2-1}\left(\frac{1}{\kappa}\right)^{2m+1}P_{l}^{(2m+1)}(1)
\end{eqnarray*}
When $\kappa \gg 1$, $\coth\kappa\approx 1$ and we have
\begin{eqnarray*}
B_{l} &\approx& \sum_{n=0}^{l}\left(-\frac{1}{\kappa }\right)^{n}P_{l}^{(n)}(1) \\
 &=&\sum_{n=0}^{l}\left(-\frac{1}{\kappa}\right)^{n}\frac{(l+n)!}{(l-n)!n!2^{n}} 
\end{eqnarray*}
Since $(l+n)!/(l-n)!\approx [l(l+1)]^n$ for $l\gg n$, we have
\begin{eqnarray}
B_{l}&\approx& \sum_{n=0}^{l}\left(-\frac{1}{\kappa}\right)^{n}\frac{[l(l+1)]^{n}}{n!2^{n}} \nonumber\\
 &\approx&\exp\left[-\frac{l(l+1)}{2\kappa}\right] \label{eq:app}
\end{eqnarray}
This is the same approximation as calculated for a plane Gaussian function
\cite{Gott1990,White1992,Tegmark1996}.

In Fig.~2, the spherical harmonics coefficients for Fisher-von Mises functions are compared to those for
plane Gaussian functions.
The coefficients for Gaussian functions were calculated numerically.
The approximation Eq.~(\ref{eq:app}) are also shown for comparison.
It is clear that the approximation represents the coefficients for Fisher-von Mises functions very well
when $\sigma\lesssim1$.
On the other hand, the Eq.~(\ref{eq:app}) approximates the coefficients of
Gaussian function only when $\sigma\ll1$.

\begin{figure}[t!]
\includegraphics[width=8.5cm]{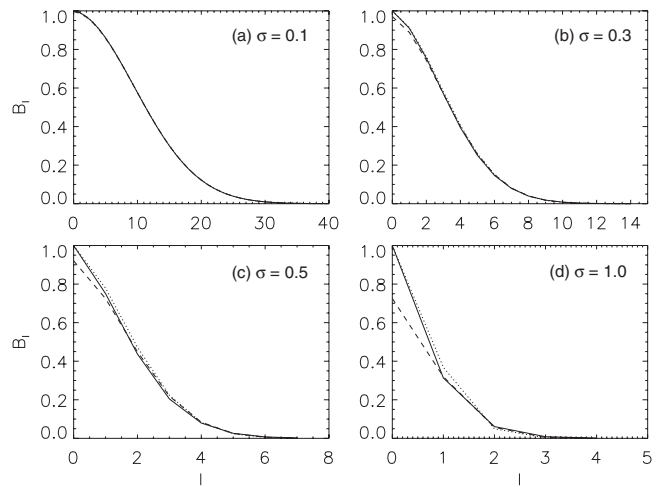}
\caption{Comparison of the spherical harmonic coefficients for Gaussian and Fisher-von Mises functions
with beam widths (a) $\sigma = 0.1$, (b) 0.3, (c) 0.5, and (d) 1.0.
Solid and dashed lines denote the coefficients calculated for Fisher-von Mises and Gaussian functions, respectively.
Dotted line represents the approximation $\exp[-l(l+1)/2\kappa]$.}
\label{fig1}
\end{figure}

\section{Generating Function}
The recursion relation Eq.~(\ref{eq:recursion}) is a difference equation,
thus we can obtain a differential equation equivalent to the
difference equation \cite{Bender1978}, by defining that
\begin{eqnarray}
F(x)=\sum_{l=0}^{\infty}\frac{B_l x^l}{l!}.
\end{eqnarray}
To obtain the differential equation for $F(x)$ we multiply Eq.~(\ref{eq:recursion}) by
$x^{l-1}/(l-1)!$ and sum from $l=1$ to $\infty$.
The first term gives
\begin{eqnarray*}
\sum_{l=1}^{\infty}B_{l+1}\frac{x^{l-1}}{(l-1)!}&=&\frac{d^2}{dx^2}\left[F(x)-B_0-B_1 x\right]\\
&=&\frac{d^2 F(x)}{dx^2}.
\end{eqnarray*}
The second term gives
\begin{eqnarray*}
\sum_{l=1}^{\infty}B_{l-1}\frac{x^{l-1}}{(l-1)!}=F(x).
\end{eqnarray*}
The third term gives
\begin{eqnarray*}
\sum_{l=1}^{\infty}(2l+1)B_l \frac{x^{l-1}}{(l-1)!}=2x\frac{d^2 F(x)}{dx^2}+3\frac{dF(x)}{dx}.
\end{eqnarray*}
Combining these three results gives the differential equation for $F(x)$
\begin{equation}
(1+2x/\kappa)F''+3F'/\kappa-F=0.
\end{equation}
This equation is simplified by substituting $y=(1+2x/\kappa)^{1/2}$,
\begin{eqnarray*}
\frac{y}{\kappa^2}\frac{d^2 F}{dy^2}+\frac{2}{\kappa^2}\frac{dF}{dy}-yF=0.
\end{eqnarray*}
Thus, we obatin
\begin{eqnarray*}
\frac{1}{\kappa^2}\frac{d^2}{dy^2}(yF)-yF=0.
\end{eqnarray*}
The general solution of the above equation is given by
\begin{eqnarray*}
yF(x)=C_1 e^{\kappa y}+C_2 e^{-\kappa y}.
\end{eqnarray*}
Using the initial conditions, $F(0)=B_0=1$ and
$F'(0)=B_1=\coth\kappa-1/\kappa$,
we obtain the generating function for the spherical coefficients $B_l$,
\begin{eqnarray}
F(x)&=&\frac{\sinh \left(\kappa\sqrt{1+2x/\kappa}\right)}{(\sinh\kappa)\sqrt{1+2x/\kappa}}.
\end{eqnarray}
The formula for $B_l$ is, then, given by
\begin{eqnarray}
B_{l} &=&\frac{d^{l}F(x)}{dx^{l}}\bigg|_{x=0}.
\end{eqnarray}
We may expand the generating function as, when $\kappa\ll 1$,
\begin{eqnarray*}
F(x) &=& \frac{1}{\sinh\kappa}\sum_{n=0}^{\infty}\frac{\kappa^{2n+1}(1+2x/\kappa)^n}{(2n+1)!}
\end{eqnarray*}
Thus, we obtain the coefficients $B_l$ for $\kappa\ll 1$
\begin{eqnarray*}
B_l&=&\frac{(2/\kappa)^l}{\sinh\kappa}
     \sum_{n=l}^{\infty}\frac{\kappa^{2n+1}}{(2n+1)!}\frac{n!}{(n-l)!}\\
     &\approx&\frac{2^l l!}{(2l+1)!}\kappa^l.
\end{eqnarray*}

\section{SUMMARY}
A recursion formula for the spherical harmonic transformation coefficients
of a Fisher-von Mises function, which is the spherical version
of a Gaussian function, was obtained.
The formula may be used for the convolution of all-sky survey maps,
such as CMB maps and FIMS FUV sky maps, to obtain smoothed sky maps.

We also investigated the mathematical properties of the spherical harmonic coefficients
and found that the coefficients approaches $\exp[-l(l+1)/2\kappa]$, and $\propto\kappa^l$
when $\kappa\gtrsim 1$ and $\kappa\ll1$, respectivley.

\appendix



\begin{references}
\bibitem{Gott1990} J. R. Gott III, C. Park, R. Juszkiewicz, W. E. Bies, D. P. Bennett,
   F. R. Bouchet, \& A. Stebbins, \apj, {\bf 352}, 1 (1990)
\bibitem{White1992} M. White, Phys.~Rev.~D {\bf 46}, 4198 (1992)
\bibitem{Fosalba2002} P. Fosalba, O. Dor{\'e}, \& F. R. Bouchet,
    Phys.\ Rev.\ D {\bf 65}, 63003 (2002)
\bibitem{Tristram2004} M. Tristram, J. F. Mac{\'i}as-P{\'e}rez, \& C. Renault, Phys.~Rev.~D {\bf 69},
   123008 (2004)
\bibitem{Tegmark1996} M. Tegmark, D. H. Hartmann, M. S. Briggs, \& C. A. Meegan, \apj, {\bf 468} (1996)
\bibitem{Finkbeiner2003} D. P. Finkbeiner, \apj{Supp. Ser.}, {\bf 146}, 407 (2003)
\bibitem{Fisher1987} N. I. Fisher, T. Lewis, \& B. J. J. Embleton,
   {\it Statistical Analysis of Spherical Data} (Cambridge: Cambridge Univ. Press, 1987), p.86
\bibitem{Seon2003} K. I. Seon, S. Park, J. H. Park, I. S. Yuk, H. Jin, U. W. Nam, W.Han, K. S. Ryu, D. H. Lee,
   S. H. Oh, Y. S. Park, E. J. Korpela, J. Edelstein, K. Nishikida, J. H. Shinn, J. G. Rhee, K. W. Min,
   and Y. H. Kim, J.~Korean Phys.~Soc.~{\bf 43}, 565 (2003).
\bibitem{Seon2005a} K. I. Seon, I. S. Yuk, U. W. Nam, J. H. Shinn, J. H. Park, D. H. Lee, I. J. Kim,
   K. S. Ryu, K. W. Min, H. Jin, W. Han, J. Edelstein, E. Korpela, \& K. Nishikida,
   J.~Korean Phys.~Soc.~{\bf 46}, 1270 (2005)
\bibitem{Seon2005b} K. I. Seon, W. Han, D. H. Lee, U. W. Nam, J. H. Park, I. S. Yuk, H. Jin, K. W. Min,
   K. S. Ryu, J. Edelstein, \& E. Korpela, J.~Korean Astron.~Soc.~{\bf 38}, 69 (2005)
\bibitem{Arfken1985} G. Arfken,
   {\it Mathematical Methods for Physicists, 3rd ed.} (Orlando: Academic Press Inc., 1985), pp.647--649
\bibitem{Bender1978} C. M. Bender, \& S. A. Orszag,
   {\it Advanced Mathematical Methods for Scientists and Engineers} (New York: McGraw-Hill Inc., 1978),
   p.46
\end{references}
\end{document}